\documentclass[twocolumn]{aastex631}
\shorttitle{Parabolic-like Trend in SiO Ratios throughout the Galactic CMZ}
\shortauthors{Takekawa et al.}

\begin{document}

\title{Parabolic-like Trend in SiO Ratios throughout the Central Molecular Zone:\\ Possible Signature of a Past Nuclear Activity in the Galactic Center}

\author[0000-0001-8147-6817]{Shunya Takekawa}
\affiliation{Department of Applied Physics, Faculty of Engineering, Kanagawa University, 3-27-1 Rokkakubashi, Kanagawa-ku, 
Yokohama, Kanagawa 221-8686, Japan}

\author[0000-0002-5566-0634]{Tomoharu Oka}
\affiliation{Department of Physics, Institute of Science and Technology, Keio University, 3-14-1 Hiyoshi, Kohoku-ku, Yokohama, Kanagawa 223-8522, Japan}
\affiliation{School of Fundamental Science and Technology, Graduate School of Science and Technology, Keio University, 3-14-1 Hiyoshi, Kohoku-ku, Yokohama, Kanagawa 223-8522, Japan}

\author[0000-0002-1663-9103]{Shiho Tsujimoto}
\affiliation{School of Fundamental Science and Technology, Graduate School of Science and Technology, Keio University, 3-14-1 Hiyoshi, Kohoku-ku, Yokohama, Kanagawa 223-8522, Japan}

\author[0000-0003-3853-1686]{Hiroki Yokozuka}
\affiliation{School of Fundamental Science and Technology, Graduate School of Science and Technology, Keio University, 3-14-1 Hiyoshi, Kohoku-ku, Yokohama, Kanagawa 223-8522, Japan}

\author[0000-0002-6824-6627]{Nanase Harada}
\affiliation{National Astronomical Observatory of Japan, 2-21-1 Osawa, Mitaka, Tokyo 181-8588, Japan}
\affiliation{Department of Astronomy, School of Science, The Graduate University for Advanced Studies (SOKENDAI), 2-21-1 Osawa, Mitaka, Tokyo, 181-1855 Japan}

\author[0000-0003-4732-8196]{Miyuki Kaneko}
\affiliation{School of Fundamental Science and Technology, Graduate School of Science and Technology, Keio University, 3-14-1 Hiyoshi, Kohoku-ku, Yokohama, Kanagawa 223-8522, Japan}

\author[0000-0003-2735-3239]{Rei Enokiya}
\affiliation{Faculty of Science and Engineering, Kyushu Sangyo University, 2-3-1 Matsukadai, Higashi-ku, Fukuoka 813-8503, Japan}
\affiliation{Department of Physics, Institute of Science and Technology, Keio University, 3-14-1 Hiyoshi, Kohoku-ku, Yokohama, Kanagawa 223-8522, Japan}

\author[0000-0002-9255-4742]{Yuhei Iwata}
\affiliation{Mizusawa VLBI Observatory, National Astronomical Observatory of Japan, 2-12 Hoshigaoka, Mizusawa, Oshu, Iwate 023-0861, Japan}



\begin{abstract}
We report the discovery of a characteristic trend in the intensity ratios of SiO emissions across the Central Molecular Zone (CMZ) of our Galaxy.
Using the Nobeyama Radio Observatory 45-m telescope, we conducted large-scale, high-sensitivity imaging observations in molecular lines including SiO {\it J}=2--1 and CS {\it J}=2--1.
By identifying SiO-emitting clouds and {examining} their intensity ratios relative to the other molecular lines, we unveiled a {parabolic-like} trend showing lower ratios near the Galactic nucleus, Sgr A$^*$, with gradual increases toward the edges of the CMZ.
This pattern suggests a {possible} outburst of the nucleus within the last $\sim 10^5$ yr, which may have propagated through the entire CMZ with strong shocks.
Alternatively, the observed trend may also be attributed to the destruction of small dust grains by high-energy photons.
Our results {can potentially lead to} a new perspective on the history of nuclear activity and its impact on the surrounding molecular environment.
\end{abstract}



\section{Introduction} \label{sec:intro}
Our Galaxy harbors a supermassive black hole (SMBH) with a mass of $4\times 10^6$ $M_\odot$ at its dynamical center, known to be a radio source Sgr~A$^*$ \citep[e.g.,][]{ghez08, gillessen09, eht22}.
{Although} the current activity of Sgr~A$^*$  is extremely quiescent  \citep[e.g.,][]{narayan98}, {particularly} compared to active galactic nuclei (AGNs), a number of observational studies suggest that it has experienced periods of energetic activity in the past.
The Fermi and eROSITA bubbles, immense $\gamma$-ray and X-ray structures extending above and below the Galactic plane, may be a remnant of an AGN activity several million years ago \citep[e.g.,][]{su10,yang22}.
On a smaller scale, the Galactic center lobe in radio wavelengths \citep{sofue84, heywood19} and the X-ray chimney \citep{ponti19} {may have originated} from similar energetic events.
Furthermore, Fe 6.4 keV fluorescent line observations of X-ray reflection nebulae suggest that Sgr~A$^*$ was considerably more luminous in X-ray a few hundred years ago \citep[e.g.,][]{koyama96, marin23}.
In addition, some authors have suggested that Sgr~A East, which is usually classified as a supernova remnant, could be formed through a tidal disruption of a star by Sgr~A$^*$ \citep{guillochon16}.
These diverse implications of episodic nuclear activities provide clues to understanding the feeding and feedback in the Galactic center.

The region within approximately 200 pc of the Galactic nucleus, known as the Central Molecular Zone (CMZ), is characterized by an enormous amount of molecular gas with high temperature ($T_{\rm k}\gtrsim 30$ K), high density ($n_{\rm H_2}\gtrsim 10^4$ cm$^{-3}$), and dominant turbulence \citep[e.g.,][]{morris96}.
The previous observations revealed the presence of widespread SiO emission in the CMZ \citep{martin-pintado97, riquelme10, jones12, tsuboi15}.
SiO is a well-established shock tracer because strong shocks can destroy dust grains, thereby sputtering silicon atoms into the gas phase, where they efficiently form SiO molecules \citep{martin-pintado92}.
Such shocks can be triggered by various phenomena, including protostellar outflows \citep[e.g.,][]{bachiller97, lu21}, supernovae \citep[e.g.,][]{dumas14, enokiya23}, cloud-cloud collisions \citep[e.g.,][]{tsuboi15, armijos-abendano20, enokiya21}, and gravitational interactions with massive compact objects \citep{oka16, takekawa19a, takekawa19b, takekawa20, kaneko23}.
The nuclear activities are another potential source of strong shocks, as suggested by the observations of NGC 1068 that AGN activity can enhance the SiO abundance in the circumnuclear region \citep{huang22}.
However, understanding exactly how the Galactic nuclear activities have affected the surrounding molecular gas {is challenging} because of the observed complexity of the CMZ clouds in their morphology, kinematics, physical conditions, and chemical composition.

Comprehensive and large-scale observations of shocked molecular gas in the CMZ are crucial for a broader understanding of the diverse phenomena {that occur} in the Galactic center.
We have conducted extensive observations throughout the CMZ using multiple 3-mm band molecular lines with the Nobeyama Radio Observatory (NRO) 45-m radio telescope.
This paper focuses {primarily} on the SiO line to systematically {identify} shocked regions.
A full data presentation along with detailed analyses will be published in a forthcoming paper.
In this paper, a distance to the Galactic center is assumed to be 8 kpc.

\section{Observations } \label{sec:obs}
We performed large-scale molecular line imaging observations {that covered} almost the entire CMZ by using the NRO 45-m telescope.
The observed region encompassed an area of $3.5\arcdeg\times 0.5\arcdeg$, ranging from Galactic longitude $l=-1.5\arcdeg$ to $+2.0\arcdeg$ and latitude $b=-0.25\arcdeg$ to $+0.25\arcdeg$.
The observed lines include millimeter lines such as SiO {\it J}=2--1, CS {\it J}=2--1, H$^{13}$CN {\it J}=1--0, H$^{13}$CO$^+$ {\it J}=1--0, and SO $N_J$=3$_2$--2$_1$.
Observations spanning $l=-1.5\arcdeg$ to $+1.5\arcdeg$ were {conducted} as part of the NRO 45-m telescope Large Program (LP187001, PI: S. Takekawa) during the period of 2019 January to May, 2020 January to April, and from 2021 January to April.
The additional observations extending from $l=+1.5\arcdeg$ to $+2.0\arcdeg$ were conducted in 2023 January and February (G22022, PI: S. Takekawa).
The total operating time of the telescope for data acquisition {was} 310 hr.

The two-sideband receiver FOREST \citep{minamidani16} and the SAM45 spectrometer \citep{kamazaki12} were utilized in the on-the-fly mode \citep{sawada08} for these observations.
Pointing errors were corrected approximately every 1.5 hr by observing the SiO maser source VX Sgr at 43 GHz {using} the H40 receiver {to ensure that} the pointing accuracy was maintained within $5\arcsec$.
The standard chopper-wheel calibration method was employed to derive the antenna temperature $T_{\rm a}^*$.
The system noise temperatures $T_{\rm sys}$ typically ranged from 150 to 350 K during the observations.
Data reduction was performed using the NOSTAR reduction package provided by NRO.
After the baseline subtraction with a linear fit, the data were resampled {using} a Bessel-Gaussian function onto a regular grid of $7.5\arcsec\times 7.5\arcsec\times 2$ km s$^{-1}$.
{To convert} the antenna temperature $T_{\rm a}^*$ to a main beam temperature $T_{\rm MB}$, {it was divided} by the main beam efficiency $\eta_{\rm MB}$=0.50 for the SiO, H$^{13}$CN, H$^{13}$CO$^+$ lines, and 0.45 for the CS and SO lines.
The RMS ($1\sigma$) noise levels of the resultant maps range approximately from 0.13 to 0.20 K.
The effective spatial resolution of the maps is $20\arcsec$, which corresponds to 0.8 pc at the Galactic center distance of 8 kpc.

\section{Results} \label{sec:results}
\begin{figure*}[htb]
\plotone{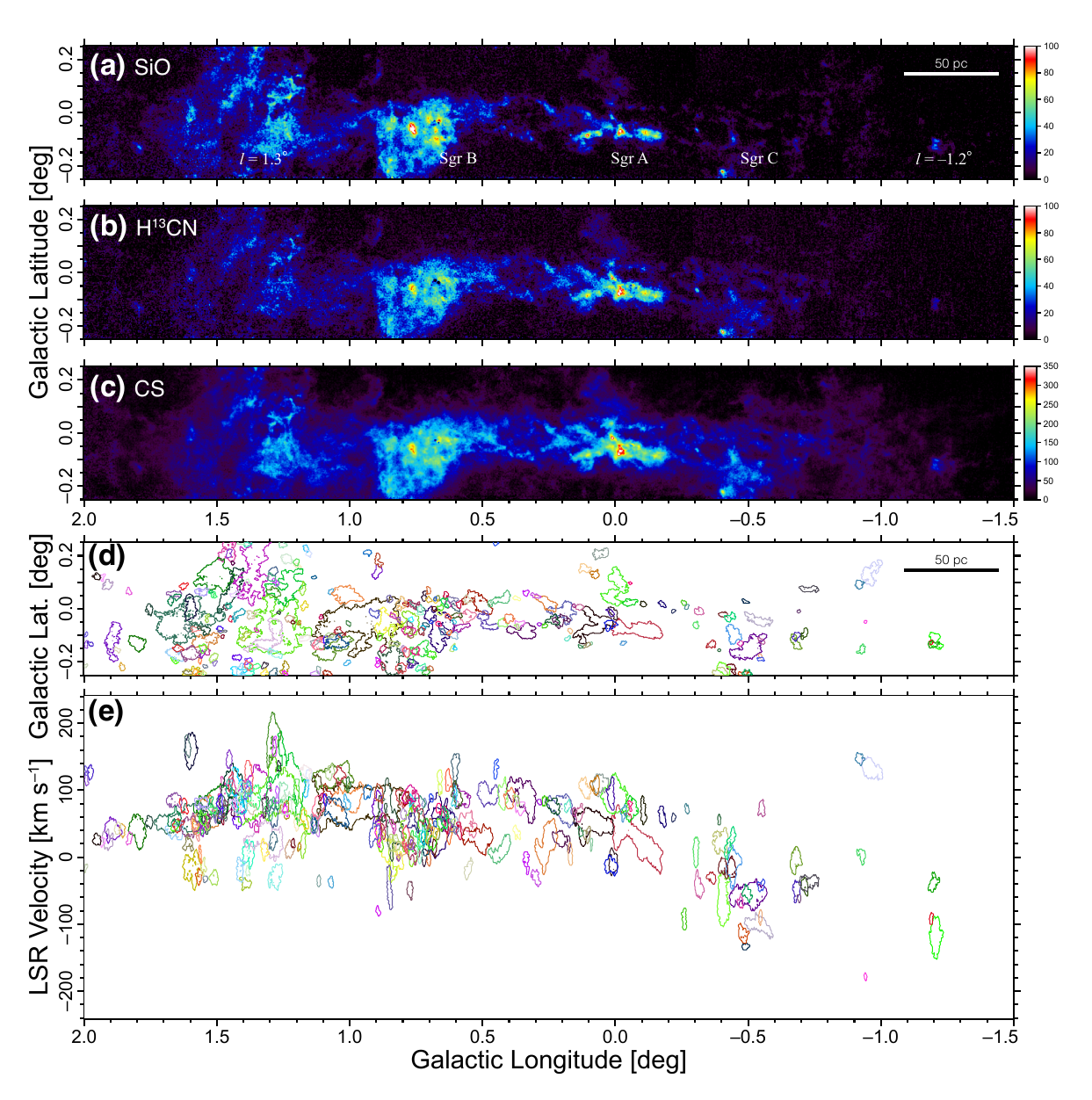}
\caption{(a)--(c) Integrated intensity maps of the SiO {\it J}=2--1 (a), H$^{13}$CN {\it J}=1--0 (b), and CS {\it J}=2--1 (c) lines.
The velocity range for the integration is $V_{\rm LSR}$=$-220$ to +220 km s$^{-1}$.
{Note that, in the region west of $l=-0.4^\circ$, H$^{13}$CO$^+$ emission was slightly detected in the velocity range $-120$ to $-100$ km s$^{-1}$, which contaminated the SiO velocity range of $+200$ to $+220$ km s$^{-1}$. Therefore, for the SiO map west of $l=-0.4^\circ$, the integration range is limited to $+200$ km s$^{-1}$.}
The unit of intensity is K km s$^{-1}$.
(d)--(e) Contours of the SiO clouds projected onto the $l$-$b$ plane (d) and the $l$-$V$ plane (e).
{Contours with the same assigned color} in both panels correspond to the same SiO cloud.
  \label{fig1}}
\end{figure*}

Figures \ref{fig1}(a)--(c) show the integrated intensity maps of the SiO {\it J}=2--1, H$^{13}$CN {\it J}=1--0, and CS {\it J}=2--1 lines, respectively.
The CS map most clearly outlines the overall spatial distribution of {the} molecular clouds in the CMZ.
The SiO and H$^{13}$CN maps exhibit more localized structures compared to the CS map.
A number of clumpy structures {are} distinctly {present} in the SiO map.
The SiO line prominently delineates not only the well-known molecular complexes of the Sgr A, Sgr B, and $l$=1.3$\arcdeg$ regions but also the $l$=$-1.2\arcdeg$ cloud \citep{tsujimoto18}.
Furthermore, {based on} our observations, a widespread distribution of faint, diffuse SiO clouds extending from the western side of the Sgr C region at $l\simeq-0.6\arcdeg$ to $-1.2\arcdeg$ {is evident}.
A full presentation of the detailed position-velocity structures, including those from the other lines, will be provided in a forthcoming paper.

\subsection{SiO cloud identification} \label{sec:3.1}
To investigate the properties of shocked molecular clouds throughout the CMZ, we performed cloud identification on the SiO map by {utilizing} the Spectral Clustering for Interstellar Molecular Emission Segmentation (SCIMES) algorithm \citep{colombo15}.
SCIMES identifies and segments significant molecular clouds in a data cube by applying a spectral clustering technique based on graph theory to the graph representation generated by the dendrogram algorithm \citep{rosolowsky08}.
The principles and methodology of SCIMES are comprehensively described in \citet{colombo15}.
The input parameters for the dendrogram were configured as follows: \texttt{min\_value} = 0.39 K ($3\sigma$), \texttt{min\_delta} = 0.39 K ($3\sigma$), and \texttt{min\_npix} = 34.
This \texttt{min\_npix} value corresponded to 6 times the number of voxels within the spatial resolution.
{After} executing SCIMES on the SiO map, a total of 258 clouds {were identified}.

Figures \ref{fig1}(d) and (e) show the contours of the identified SiO clouds projected onto the $l$-$b$ and  $l$-$V$ planes, respectively.
{Although} many SiO clouds have been successfully identified, not all structures identified by SCIMES necessarily represent actual individual molecular clouds.
The diameter ($d$) of the SiO clouds predominantly ranged from 1 to 5 pc, and the velocity dispersion ($\sigma_{V}$) {was} from 2 to 10 km s$^{-1}$.
Here, the diameter $d$ is defined as $d=2\sqrt{A/\pi}$, where $A$ is the area of each cloud's contour projected onto the $l$-$b$ plane.

More than 80\% of the clouds, 218 out of 258, are located at positive Galactic longitudes, highlighting the remarkable asymmetry of the CMZ.
The eastern side from $l$=0.6$\arcdeg$, encompassing the Sgr B and $l$=1.3$\arcdeg$ complexes, is particularly crowded with SiO clouds, containing approximately 70\% of the total within this range. 
Conversely, on the western side beyond the Sgr C region, which is an area not covered in the previous SiO surveys by \citet{jones12} and \citet{tsuboi15}, SiO clouds are fewer and more sparsely distributed. 
This region includes notable clouds such as the $l$=$-1.2\arcdeg$ cloud \citep{tsujimoto18} and the Pigtail cloud \citep{matsumura12}.

\begin{figure*}[tbh]
\plotone{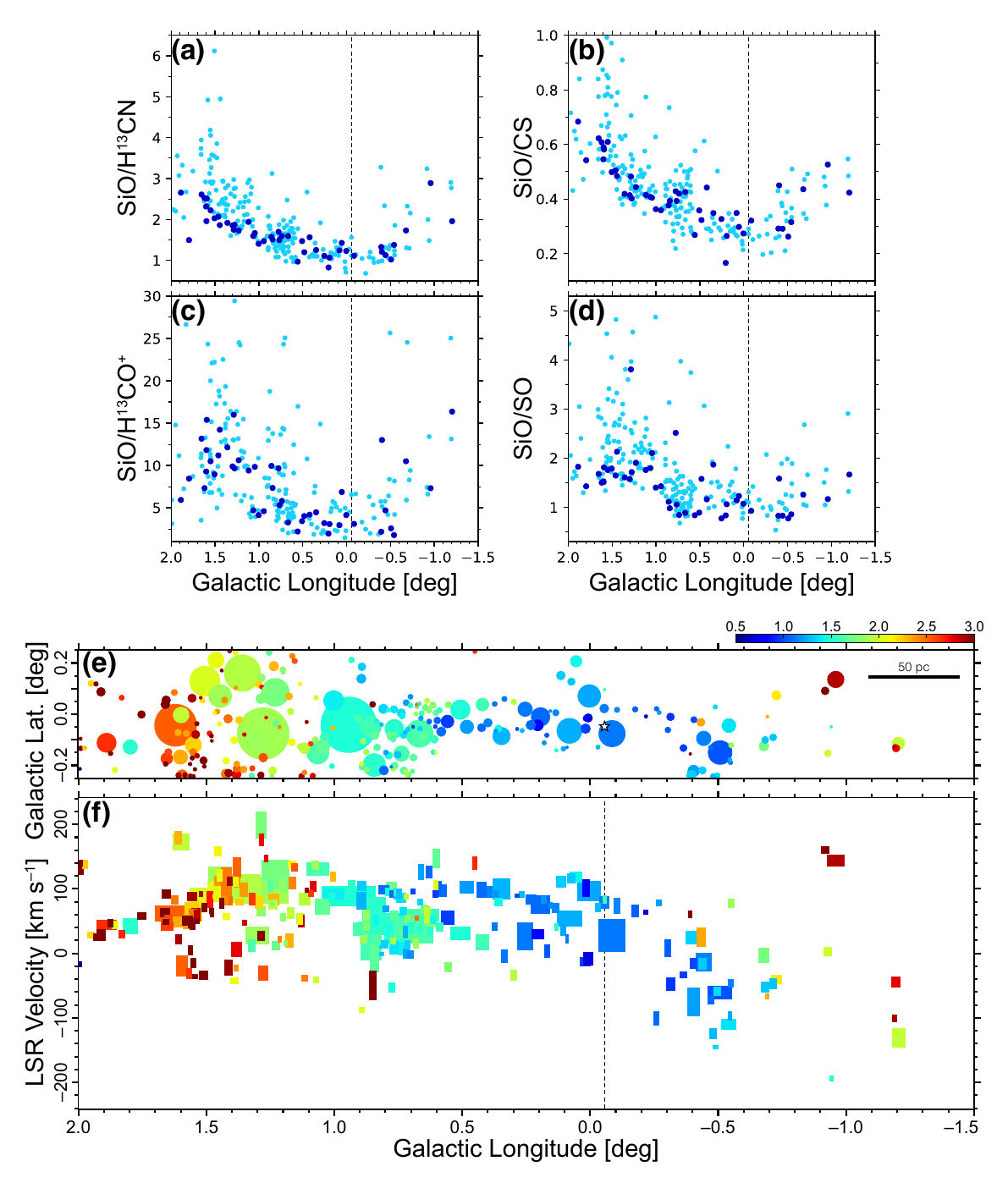}
\caption{
(a)--(d) SiO intensity ratios of SiO clouds as a function of the Galactic longitude.
Each data point represents an individual SiO cloud.
Clouds with diameters $d$ larger than 5 pc are shown in dark blue.
Panel (a) displays the ratio of SiO/H$^{13}$CN, (b) SiO/CS, (c) SiO/H$^{13}$CO$^{+}$ and (d) SiO/SO.
(e)--(f) Spatial and velocity distributions of SiO clouds {with colors assigned according to} the SiO/H$^{13}$CN ratio.
In the $l$-$b$ plane (e), the diameter of each circle corresponds to the diameter $d$ of the SiO cloud.
In the $l$-$V$ plane (f), the width and height of each rectangle represent the cloud's diameter $d$ and velocity width, respectively.
The velocity width is defined as $2\sqrt{2\ln{2}}\sigma_{V}$.
The star {symbol} in panel (e) and vertical dashed lines in the other panels indicate the position of Sgr A$^*$. 
  \label{fig2}}
\end{figure*}

\subsection{Parabolic{-like} trend in the SiO intensity ratios}
{As} strong shocks can enhance the abundance of SiO in the interstellar medium (ISM) through the sputtering of dust grains, the intensity ratios of SiO lines {relative} to those of other molecules serve as useful probes for shocked molecular gas.
For each SiO cloud, we measured the intensity ratios of the SiO line {relative} to those of the other lines.
The ratio for a cloud was determined by calculating the ratio between the integrated intensities over all its voxels.
{By analyzing the results}, we found an interesting correlation between these ratios and the spatial distribution of the SiO clouds.
Figures \ref{fig2}(a)--(d) show the plots of the Galactic longitude position versus the ratios of SiO/H$^{13}$CN, SiO/CS, SiO/H$^{13}$CO$^{+}$, and SiO/SO, respectively.
A common trend {is evident for} all four ratios: the SiO ratio has a minimum in the vicinity of the Galactic nucleus and gradually increases toward the edges of the CMZ.
The ratios {have a} peak {near} $l\simeq1.6\arcdeg$ on the eastern side, {whereas} on the western side, they appear to {increase} until $l\simeq-1.2\arcdeg$, although the local maximum point remains unclear.

Several ratios notably deviate upward from the {parabolic}-like trend.
This tendency {is prominent} in smaller ($d < 5$ pc) clouds, and the deviation {is} significantly smaller when only larger clouds ($d > 5$ pc; dark blue in Figures \ref{fig2}(a)--(d)) {are considered}, which follows the trend relatively well.
The deviations of the ratios are particularly pronounced {near} the peak point ($l\simeq1.6\arcdeg$) and the Sgr B region ($l\simeq0.7\arcdeg$).
The significant scatter of the H$^{13}$CO$^{+}$ ratio is attributed to faint H$^{13}$CO$^{+}$ emissions. 
Figures \ref{fig2}(e) and (f) depict the spatial and velocity distributions of SiO/H$^{13}$CN, respectively.
They also highlight the gradual change in the SiO ratio with increasing projected distance from the nucleus.
In addition, the SiO clouds associated with the velocity components of the expanding molecular ring \citep[EMR;][]{kaifu72, scoville72} {exhibited} high SiO ratios {near} $(l,\ V_{\rm LSR})\simeq(-0.95\arcdeg,\ 150\ \rm km\ s^{-1})$ and $(1.5\arcdeg,\ -30\ \rm km\ s^{-1})$ as shown in Figure \ref{fig2}(f).
Although SiO enhancements have previously been reported in the Sgr B, $l$=1.3$\arcdeg$, and 1.6$\arcdeg$ complexes \citep{martin-pintado97, huttemeister98, tanaka07, menten09, riquelme13, tsujimoto21, busch22}, this is the first time such a large-scale SiO trend is seen across the CMZ.
The local variation in these ratios indicates the diversity in the physical and chemical conditions of individual SiO clouds, {whereas} the overall parabolic{-like} trend likely reflects an intrinsic property of the CMZ.

\section{Discussions} \label{sec:discussion}
\subsection{Origin of the parabolic{-like} trend in the SiO ratios} 
The common trend in the line intensity ratios of SiO {relative} to the other lines suggests a distinctive variation in SiO abundance.
SiO abundance can be enhanced {via} strong shocks induced by various sources such as interactions with supernovae \citep[e.g.,][]{dumas14}, protostellar outflows \citep[e.g.,][]{bachiller97}, and cloud-cloud collisions \citep[e.g.,][]{armijos-abendano20}.
Frequent cloud-cloud collisions are expected at sites of gas inflow into the CMZ \citep{sormani19} and at intersections between the $x_1$ and $x_2$ orbits \citep{binney91}.
Indeed, in the candidate sites such as the $l$=1.3$\arcdeg$ complex \citep{busch22} and the Sgr B complex \citep{molinari11, enokiya22}, SiO clouds are densely populated with high SiO ratios.
However, considering that molecular clouds within the CMZ can travel only {approximately} 10 pc in the SiO depletion timescale of $\sim10^5$ yr, it is unlikely that these orbital interactions alone could {account for} the {observed} gradual trend in the SiO ratios across the CMZ spanning 400 pc.
Moreover, a single supernova explosion would also be insufficient to disturb the entire CMZ, and protostellar outflows are even more localized phenomena.
{Although} the aforementioned factors should contribute to the local enhancements in the SiO ratios, the overall parabolic{-like} trend is more likely attributed to a larger scale shock arising from a different cause.

\subsection{Past outburst of Sgr A$^*$?}
\begin{figure}[tb]
\begin{center}
\includegraphics[width = 8.5 cm]{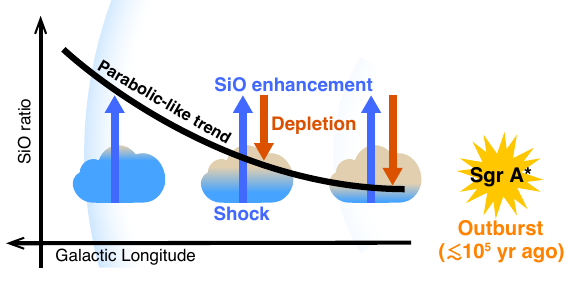}
\caption{
 Schematic of the nuclear outburst scenario. 
 Fast wind (or ejecta) driven by a past nuclear activity within the last $\sim10^5$ yr sequentially induced the SiO enhancement {via} strong shocks.
 SiO {was depleted} over time after the shock dissipated, {thereby} reducing the temporarily enhanced SiO.
 Thus, a parabolic{-like} SiO trend emerges.
  \label{fig3}}
  \end{center}
\end{figure}

The symmetrical nature of the parabolic{-like} trend with respect to the Galactic nucleus, Sgr~A$^*$, implies that it may originate from the nucleus itself or its immediate vicinity.
Hence, we propose that a past explosive event associated with Sgr~A$^*$ may be responsible for this parabolic{-like} trend.
A schematic of this scenario is illustrated in Figure \ref{fig3}.
In our scenario, an outburst from Sgr~A$^*$ within the last $\sim10^5$ yr is hypothesized to {have driven} fast wind or ejecta propagating throughout the CMZ  with a velocity of $\sim$2000 km s$^{-1}$ ($\sim$200 pc/$10^5$ yr).
This outburst sequentially impacted molecular clouds closer to Sgr~A$^*$, inducing strong shocks and consequently enhancing their SiO abundance.
Over time, as the shock wave dissipated, the SiO abundance would decrease owing to the depletion onto dust grains, leading to lower SiO ratios.
Therefore, clouds near Sgr~A$^*$ exhibit relatively lower SiO ratios because of the longer time elapsed since the outburst, while more distant {clouds that were more recently affected by the outburst} show higher ratios, resulting in the parabolic{-like} trend.

The peak of the SiO ratios {near} $l$=$1.6\arcdeg$ may signify the current forefront of this blast wave.
Note that the timescale of $10^5$ yr is constrained by the SiO depletion timescale of $\tau_{\rm dep}\sim 10^9/(n_{\rm H_2}/\rm cm^{-3})\rm \ yr$ \citep[e.g.,][]{martin-pintado92} for a typical CMZ cloud density of $n_{\rm H_2}\sim 10^4$ cm$^{-3}$ \citep[e.g.,][]{tanaka18}.

\begin{figure*}[!htb]
\begin{center}
\includegraphics[width = 16 cm]{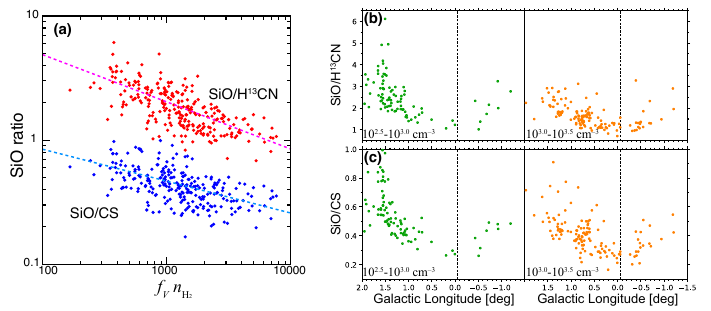}
\caption{
 (a) SiO intensity ratios of SiO clouds as a function of their average density.
 Red and blue dots indicate the SiO/H$^{13}$CN and SiO/CS ratios, respectively. The best-fit lines are {represented as dashed lines}.
 {(b)--(c) SiO/H$^{13}$CN and SiO/CS intensity ratios of the SiO clouds within the restricted average density ranges as a function of the Galactic longitude, respectively. Green points are the clouds with average {densities} from 10$^{2.5}$ to $10^{3.0}$  cm$^{-3}$, and orange points {have average densities} from 10$^{3.0}$ to $10^{3.5}$  cm$^{-3}$.
 The vertical dashed lines indicate the position of Sgr A$^*$.}
  \label{fig4}}
  \end{center}
\end{figure*}

{As} the SiO depletion timescale is inversely proportional to the cloud density, for the depletion effect to cause {a} parabolic{-like} trend in {the} SiO ratios, clouds with higher ratios must have densities comparable to or lower than those with lower ratios.
We estimated the densities of SiO clouds from the H$^{13}$CN line emission, assuming local thermodynamic equilibrium (LTE) and optically thin conditions.
The LTE masses were calculated with an excitation temperature of 5 K and a fractional abundance of [H$^{13}$CN]/[H$_2$]=$4 \times 10^{-10}$ \citep{goicoechea18}.
{The} excitation temperature was inferred from the HCN/H$^{13}$CN ratios, {which} typically ranged between 3 and 7 K with a median of 5 K.
We then derived the density by dividing the mass by {the corresponding} volume, which was {calculated} as $\frac{4\pi}{3}(\frac{d}{2})^3$.
The {derived density is} the average value denoted as $f_V n_{\rm H_2}$, where $f_V $ is a volume-filling factor.
Figure \ref{fig4}(a) shows the SiO ratios of SiO/CS and SiO/H$^{13}$CN plotted against the average density {values}.
{These values are} significantly lower by an order of magnitude {compared to} the typical value of $10^4$ cm$^{-3}$ in the CMZ, {which may be indicative of} low volume filling factors ($f_V\sim 0.1$).
{Distinct} negative correlations can be confirmed: clouds with higher SiO ratios tend to have lower densities, implying longer SiO depletion timescales.
{This result {satisfies} the necessary conditions without conflicting with the nuclear outburst scenario.}

{
If the parabolic{-like} trend reflects the time evolution following SiO enhancements in order of proximity to the nucleus, the trend should also emerge when restricted to clouds with similar densities.
Figures \ref{fig4}(b) and (c) display the SiO/H$^{13}$CN and SiO/CS ratio plots similar to Figures \ref{fig2}(a) and (b), but confined to the average density ranges of 10$^{2.5}$--10$^{3.0}$ (green) and 10$^{3.0}$--10$^{3.5}$ (orange), respectively.
A similar trend within each density range may suggest that the trend reflects not only the difference in SiO depletion timescales but also the time elapsed since the SiO enhancements.
}

If our scenario is correct, the trend may also be present along the Galactic latitude.
Figure \ref{fig2}(e) suggests a slight increase in the SiO ratio from Sgr~A$^*$ toward the northern side of the polar arc at $(l,\ b)\simeq (0.1\arcdeg,\ 0.2\arcdeg)$, which could be another support.
{At the higher Galactic latitude above Sgr C, there is a warm dust region, AFGL 5376, associated with highly turbulent molecular gas, which suggests that it may be influenced by strong shocks \citep{uchida94, ponti21}.
Although the connection between this shocked region and the outburst is currently unclear, future investigations of the SiO trends and detailed physical conditions of such molecular gas away from the Galactic plane could provide further evidence for the outburst.}
Additionally, the SiO clouds detected in the velocity components of the EMR exhibited high SiO ratios, {near} $(l,\ V_{\rm LSR})\simeq(-0.95\arcdeg,\ 150\ \rm km\ s^{-1})$ and $(1.5\arcdeg,\ -30\ \rm km\ s^{-1})$ (Figure \ref{fig2}(f)).
While the large-scale expansion of the CMZ remains controversial \citep[e.g.,][]{oka20, sofue22, henshaw23}, the possible outburst may impact the EMR, causing strong shocks.

\subsection{Radiative destruction of dust grains}
{
Another possible mechanism for the SiO enhancement is the destruction of small dust grains by high-energy photons, rather than by mechanical shocks. 
The time required to illuminate the entire CMZ is approximately 700 yr, significantly shorter than the typical SiO depletion timescale of $10^5$ yr.
This could imply that, in the radiative case, the SiO trend predominantly reflects the density distribution of the clouds.
However, the parabolic{-like} trend was also confirmed in the clouds with similar densities (Figures \ref{fig4}(b) and (c)).
This {may indicate} that the actual depletion timescale of SiO could be unexpectedly short ($\lesssim 10^3$ yr), or perhaps reflect a radial distribution of dust properties such as grain sizes and charges. 
{Although} the radiation-induced enhancement implies an abundant presence of nano-sized silicate grains throughout the CMZ \citep{amo-baladron09}, direct detection of silicate nanoparticles in the ISM has yet to be achieved, potentially enabled by future sensitive infrared spectroscopy \citep[e.g.,][]{zeegers23}.

At this point, {conclusively  determining} the cause of the parabolic{-like} trend {is difficult}.
Further observational and theoretical studies, including more precise measurement of SiO abundance, quantification of dust properties, and chemical modeling, are necessary to fully understand the underlying mechanisms responsible for the parabolic{-like} trend, as well as {comparison} with other nearby galaxies.
}

\section{Conclusions} \label{sec:conclusion}
We conducted the large-scale imaging observations with the NRO 45 m telescope, covering almost the entire CMZ in molecular lines including SiO {\it J}=2--1.
The obtained maps are the most detailed and sensitive images of the CMZ ever achieved {using} single dishes.
To reveal shocked regions, we identified SiO clouds and examined their intensity ratios of SiO over the H$^{13}$CN, CS, H$^{13}$CO$^+$ and SO lines.
We discovered a characteristic {parabolic}-like trend in these SiO ratios relative to the Galactic longitude positions of the clouds.
This trend exhibited a minimum in the vicinity of Sgr~A$^*$ and gradually increased toward the edges of the CMZ.
We proposed that the parabolic{-like} trend can possibly be explained by a past outburst from Sgr~A$^*$ that occurred within the last $\sim10^5$ yr.
To definitively determine the cause of the SiO trend, it is necessary to examine the molecular and dust properties of the SiO clouds in more detail, {considering} factors other than mechanical shocks, such as the radiative destruction of small dust grains by high-energy photons.
Our study not only provides useful data sets for the CMZ but also {presents} a {potentially} new perspective on the past activity of the Galactic nucleus.





\begin{acknowledgments}
We are grateful to the staff of the Nobeyama Radio Observatory (NRO) for their outstanding support during the 45 m telescope observations.
The NRO is a branch of the National Astronomical Observatory of Japan (NAOJ), National Institutes of Natural Sciences.
This study was supported by a JSPS Grant-in-Aid for Early-Career Scientists grant No. JP19K14768, a JSPS Grant-in-Aid for Scientific Research (A) No. 20H00178, and a University Research Support Grant from the NAOJ.
We are also thankful to the anonymous referee for helpful comments and suggestions that improved this Letter.

\end{acknowledgments}

%

\vspace{5mm}
\facility{No:45m} %


\software{Astropy \citep{astropy13, astropy18, astropy22}, Matplotlib \citep{hunter07}, NumPy \citep{vanderWalt11, harris20}, SciPy \citep{ virtanen20}}








\end{document}